# Covalency a Pathway for Achieving High Magnetisation in TMFe$_2$O$_4$ Compounds

M. H. N. Assadi[1,*] and H. Katayama-Yoshida[2]

[1]*Center for Green Research on Energy and Environmental Materials (GREEN), National Institute for Materials Science (NIMS), 1-1 Namiki, Tsukuba, Ibaraki 305-0044, Japan.*

[2]*Center for Spintronics Research Network, The University of Tokyo, Bunkyo-ku, Tokyo 113-8656, Japan.*

The interplay between covalency and magnetism is non-trivial and can be harnessed for designing new functional magnetic materials. Based on a survey using density functional calculations, we show that TM–O bond covalency can increase the total magnetic moment of spinel compounds of TMFe$_2$O$_4$ composition (TM = V-Ni, Nb-Pd) which are isomorphic to the much-researched magnetite. Accordingly, PdFe$_2$O$_4$ was found to exhibit the highest magnetic moment of 7.809 $\mu_B$ per formula unit which is approximately twice that of Fe$_3$O$_4$ with $T_c$ predicted to be well above ambient. We further propose a practical method for synthesising PdFe$_2$O$_4$.

## 1. Introduction

Fe$_3$O$_4$, first attracted attention as a permanent magnet in 1500 BC.[1,2] As demonstrated in Figure 1, in Fe$_3$O$_4$, one-third of the Fe ions are tetrahedrally coordinated by O (A site) while the rest of Fe ions are octahedrally coordinated (B site). All A site Fe ions have oxidation state +3 while the B site Fe ions are equally split between +2 and +3 oxidation state. The electrical conductivity above Verwey transition temperature is achieved by electrons hopping from one Fe$^{2+}$ to an adjacent Fe$^{3+}$.[3] The mixed valency and an Fe–O–Fe angle close to 90° between the octahedral Fe ions facilitate a ferromagnetic double exchange interaction among Fe ions in B site. While a wider Fe–O–Fe angle between an A site Fe and another B site Fe, on the other hand, facilitates an antiferromagnetic superexchange interaction that aligns the spin of A site Fe ions antiparallel to the those in B site.[4] As a result, Fe$_3$O$_4$ is strongly ferrimagnetic with a remarkably high Curie Temperature of 860 K. Furthermore, since all Fe ions are in high spin states, despite the partial spin cancellation of Fe on tetrahedral sites against those of octahedral sites, Fe$_3$O$_4$ still achieves a high magnetic moment of ~4 $\mu_B$/f.u. and a room temperature magnetisation of 480 kA m$^{-1}$ making Fe$_3$O$_4$ a common permanent magnet.[5]

From a crystallographic viewpoint, Fe$_3$O$_4$, at room temperature, can be considered a member of the wider family of TMFe$_2$O$_4$ compounds in which Fe occupies the A site. That is because the crystal symmetry of the inverse spinel structure ($Fd\bar{3}m$) is preserved if the A site is swapped for an element other than Fe. Now, one wonders if the A site Fe ion is entirely substituted with another transition metal (TM) ion with smaller magnetisation, can a higher magnetic moment be obtained? If so, would the ferrimagnetism in the new compound be as robust as that of Fe$_3$O$_4$? Smaller magnetisation on A site, in principle, can be achieved by three means, a smaller magnetic moment on the substituting TM ion, stabilisation in the low spin state for the A site TM ion and more importantly by higher covalency in the TM–O bond. The large overlap between TM d and O 2p orbitals caused by strong covalent TM–O bonding can defy Hund's rule and result in a substantially smaller magnetic moment for the TM ions.[6] To find if such compound exists, we carried out a systematic theoretical investigation of possible TMFe$_2$O$_4$ compounds in which the TM ion is either a 3d (V-Ni) or a 4d (Nb-Pd) ion.

## 2. Computational Settings

Spin-polarised density functional theory (DFT) calculations were carried out using projector augmented wave method as implemented in VASP.[7] The energy cut-off was set to 550 eV, while a *k*-point mesh was produced by Monkhorst-Pack scheme with a spacing of ~0.025 Å$^{-1}$. The applied Hubbard term ($U_{\text{eff}}$) was 3 eV for all 3d TM ions and 2 eV for all 4d TM ions.[8] These $U_{eff}$ values reproduce the measured magnetic ordering and the electronic structure for $Fe_3O_4$ and $MoFe_2O_4$.[2, 9] Lattice parameters, internal atomic coordinates and magnetic moment of all compounds were allowed to relax fully to an energy threshold of $10^{-5}$ eV. Geometry optimisation was repeated with larger 2 × 2 × 2 supercell to detect any symmetry lowering distortions. Total energies were also examined with respect to the TM's high spin and low spin states. The optimised lattice parameters and important structural features are presented in Table 1 while the electronic configurations and magnetic characteristics are presented in Table 2. Electronic population localised at the ionic centres was analysed using Bader Charge Analysis Code.[10] The ionicity of the bonds was examined by calculating the electronic localisation function ($\eta$).[11] $\eta$ offers a straightforward topological analysis of bonding character.[12] The charge distribution in bonds of different compounds was characterised by examining the relative charge profile $\rho/\rho_{MAX}$ along the bond which is obtained by dividing the charge profile by the maximum charge value along a bond. $\rho/\rho_{MAX}$ demonstrates where the charge is heavily concentrated along the bond revealing the strength and type of the bonds. The strength of the magnetic interaction was examined by calculating $\Delta E$ defined as:

$$\Delta E = E_{FM}^{Total}(TMFe_2O_4) - E_{FiM}^{Total}(TMFe_2O_4) \quad \text{Eq. 1}$$

in which $E_{FM}^{Total}$ and $E_{FiM}^{Total}$ denote the total energy corresponding to the ferromagnetic and ferrimagnetic states respectively. In the ferromagnetic state, the spins of all TM ions are aligned parallel while in the ferrimagnetic state, the spin of the TM ions is aligned antiparallel to the spin of Fe ions. The larger $\Delta E$ is, the stronger the ferrimagnetic coupling is.

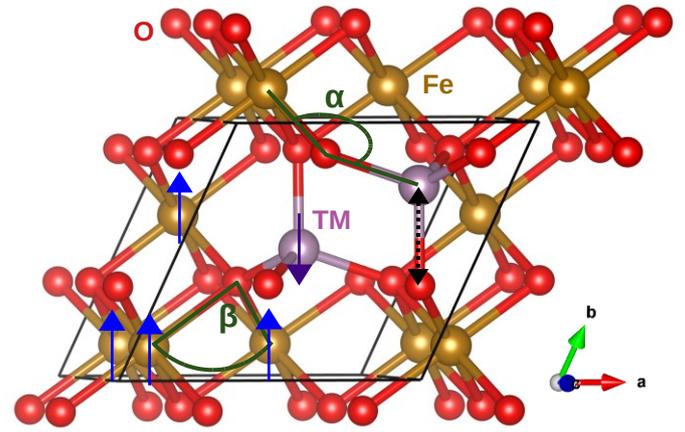

Figure 1. The primitive cell of the $Fd\bar{3}m$ structure. The arrows indicate the spin direction borne on the metal ions.

### 3. Results and Discussion

According to Table 1, all compounds except $Fe_3O_4$ and $TcFe_2O_4$ relaxed to ideal inverse spinel structure. $Fe_3O_4$, during geometry optimisation, relaxed to a monoclinic structure rather than a spinel structure. The deviation of the calculated monoclinic structure from perfect inverse spinel was however rather very minute. This deviation was mainly driven by a ~0.2% contraction in the $c$ axis of $Fe_3O_4$'s of conventional unitcell of and a ~0.5° tilt in $\beta$ accompanied by accommodating distortions in $FeO_6$ and $FeO_4$ polyhedra. The monoclinic distortion resulted in the lower total energy of 0.452 eV/f.u. in $Fe_3O_4$. Deviation from ideal cubic symmetry below Verwey transition, for long, had been controversially debated[13] but finally, it has been experimentally demonstrated that the low-temperature structure of $Fe_3O_4$ has a lower symmetry monoclinic structure below Verwey transition at 125 K.[14]

Table 1. The lattice parameters, symmetry group and the TM–O–Fe and Fe–O–Fe angles that facilitate the magnetic exchange interactions, as marked in Figure 1, of the optimised $TMFe_2O_4$ compounds. For $Fe_3O_4$ and $TcFe_2O_4$, angles were averaged. The mean deviation was smaller than 1°.

| Compound | Symmetry group | Lattice Parameters (Å) | TM–O–Fe | Fe–O–Fe |
|---|---|---|---|---|
| $VFe_2O_4$ | $Fd\bar{3}m$ | $a$ = 8.558 | 123.62° | 92.30° |
| $CrFe_2O_4$ | $Fd\bar{3}m$ | $a$ = 8.490 | 125.51° | 89.65° |
| $MnFe_2O_4$ | $Fd\bar{3}m$ | $a$ = 8.596 | 121.59° | 95.08° |
| $Fe_3O_4$ | $P121$ | $a$ = 8.469, $b$ = 8.491, $c$ = 8.467, ($\beta$ = 90.52°) | 123.46° | 92.52° |
| $CoFe_2O_4$ | $Fd\bar{3}m$ | $a$ = 8.423 | 122.65° | 93.63° |
| $NiFe_2O_4$ | $Fd\bar{3}m$ | $a$ = 8.441 | 122.71° | 93.56° |
| $NbFe_2O_4$ | $Fd\bar{3}m$ | $a$ = 8.768 | 124.70° | 91.99° |
| $MoFe_2O_4$ | $Fd\bar{3}m$ | $a$ = 8.716 | 124.12° | 91.61° |
| $TcFe_2O_4$ | $P4_122$ | $a, b$ = 6.115, $c$ = 8.648 | 123.88° | 91.93° |
| $RuFe_2O_4$ | $Fd\bar{3}m$ | $a$ = 8.697 | 121.89° | 94.66° |
| $RhFe_2O_4$ | $Fd\bar{3}m$ | $a$ = 8.701 | 120.43° | 96.61° |
| $PdFe_2O_4$ | $Fd\bar{3}m$ | $a$ = 8.709 | 120.00° | 97.21° |

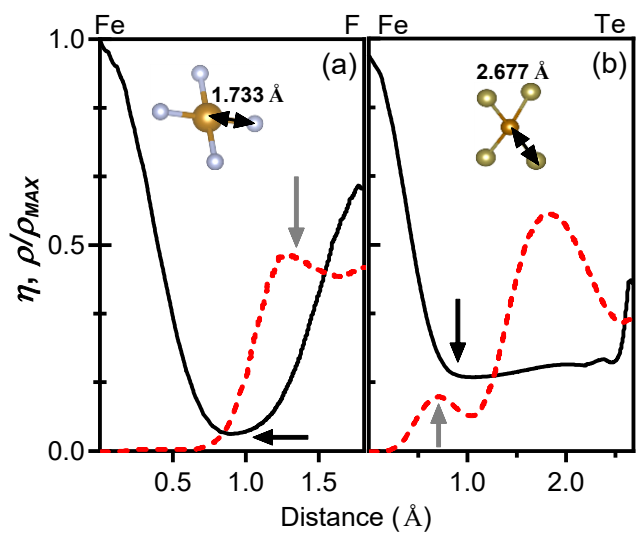

Figure 2. Electronic localisation function ($\eta$) and the relative charge ($\rho/\rho_{MAX}$) of the highly ionic FeF$_4$ and partially ionic FeTe. The dashed and solid lines denote $\eta$ and $\rho/\rho_{MAX}$ respectively. Fe in both FeF$_4$ and FeTe compounds is tetrahedrally coordinated.

The nature of the bond between the transition metal and oxygen is generally considered to be ionic on account of the ~1.6 difference in the electronegativity of the TM and O ions. Assigning ionic character to TM–O bonds can nonetheless be challenged as the electronegativity difference between TM an O is occasionally smaller than ~1.6 especially for Ni and most 4d TM ions. Consequently, in some TM oxides, the TM–O bond deviates from perfect ionicity. To obtain a tangible method for comparing the bond ionicity in TMFe$_2$O$_4$ compounds, we take a detour to establish a methodology to characterise the ionicity of Fe containing bonds. Here, we examine the $\eta$ and the $\rho/\rho_{MAX}$ profiles of FeF$_4$ and FeTe compounds which represent the extremes of ionicity and covalency respectively. The large difference in the electronegativity of Fe and F which is 2.15 confidently implies that FeF$_4$ is highly ionic while the meagre difference in the electronegativity of Fe and Te which is mere 0.27 implies that Fe–Te bond is substantially less ionic than the Fe–F bond.

According to Figure 2(a), high ionicity in FeF$_4$ manifests in a valley in the charge distribution in mid-bond area (marked with a black arrow) that drops to $\rho/\rho_{MAX} = 0.015$ and a single $\eta$ peak of ~0.5 near the anionic center (marked by a gray arrow) while $\eta$ is zero near cationic Fe centre, indicating lack of any significant orbital overlap along the bond. In the less ionic FeTe compound, as shown in Figure 2(b), $\eta$ profile is noticeably different from that of FeF$_4$ as $\eta$ shows a second peak of 0.200 at the mid-bond section (marked with a gray arrow). Furthermore, the $\rho/\rho_{MAX}$ valley has a value of 0.179, more than ten times that of Fe–F bond. FeTe's second $\eta$ peak and the more substantial $\rho/\rho_{MAX}$ at the mid-bond region is a principal feature of covalent bonding demonstrating the high probability of Fe 3d electrons appearing in the mid-bond region and thus significantly overlapping with O 2p orbitals.[15] The deviation from ionicity towards covalency reduces the magnetic moment of TM ions by cancelling the magnetic moment of the overlapped portion of d electrons.[16] In FeF$_4$, Fe's nominal oxidation state is +4, and therefore under tetrahedral crystal field, high spin Fe$^{4+}$ adopts $e^2 t_2^2$ configuration with a nominal magnetisation of 4 $\mu_B$. The calculated Fe magnetisation in FeF$_4$ is 3.444 $\mu_B$ which is close to the nominal value. In the less ionic TeFe, high spin Fe$^{2+}$ under tetrahedral crystal field nominally adopts the $e^3 t_2^3$ which should, in principle, manifest in a magnetisation of 4 $\mu_B$ for Fe. The calculated Fe magnetisation in FeTe is, however, substantially smaller at 2.90 $\mu_B$.

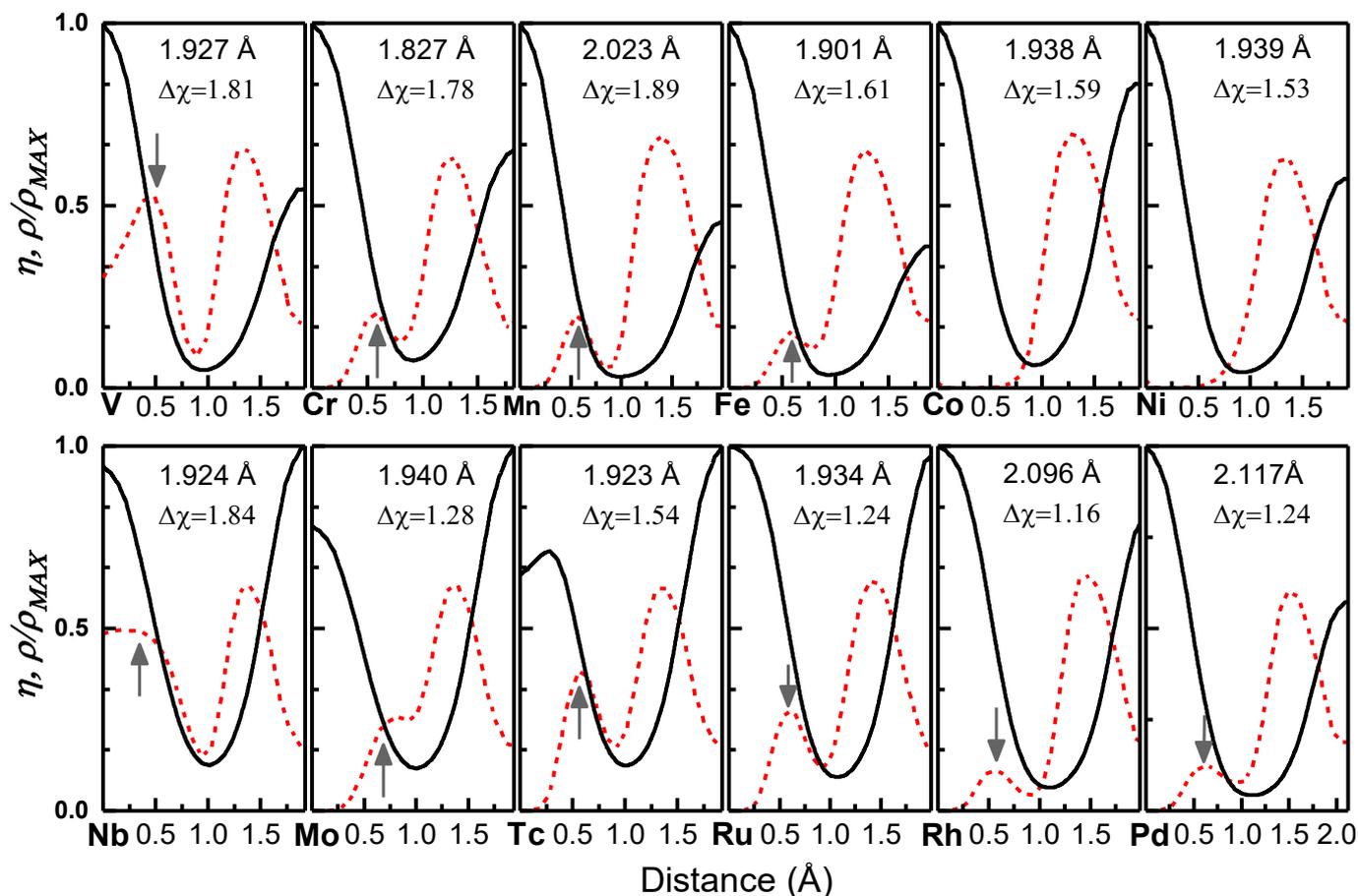

Figure 3. The $\eta$ (dashed lines) and the $\rho/\rho_{MAX}$ (solid lines) profiles of the TMFe$_2$O$_4$ compounds along the TM–O bonds. The left side of each graph coincides on the TM ion side of the bond. The values presented inside the panels are the TM–O bond length (marked with a dotted line in Figure 1) and the electronegativity difference ($\Delta X$) between TM and O.

Table 2. The energy difference between ferromagnetic and antiferromagnetic spin alignment ($\Delta E$) and the calculated magnetic moment localised on the TM ion and the nominal electronic configuration corresponding to the TM ion in tetrahedral coordination.

| Compound | $\Delta E$ (eV/f.u.) | TM Magnetisation ($\mu_B$) | TM's Electronic configuration | Compound's Mag. moment ($\mu_B$/f.u.) |
|---|---|---|---|---|
| VFe$_2$O$_4$ | 0.384 | 0.801 | $e^2 t_2^0$ | 6.931 |
| CrFe$_2$O$_4$ | 0.218 | 1.862 | $e^2 t_2^1$ | 5.863 |
| MnFe$_2$O$_4$ | 1.126 | 3.625 | $e^2 t_2^3$ | 4.981 |
| Fe$_3$O$_4$ | 0.510 | 3.960 | $e^2 t_2^3$ | 3.992 |
| CoFe$_2$O$_4$ | 0.268 | 2.667 | $e^3 t_2^3$ | 4.968 |
| NiFe$_2$O$_4$ | 1.003 | 1.553 | $e^4 t_2^3$ | 7.307 |
| NbFe$_2$O$_4$ | 0.349 | 0.374 | $e^2 t_2^0$ | 6.916 |
| MoFe$_2$O$_4$ | 0.411 | 1.486 | $e^2 t_2^1$ | 5.888 |
| TcFe$_2$O$_4$ | 0.124 | 2.221 | $e^2 t_2^2$ | 5.070 |
| RuFe$_2$O$_4$ | 0.975 | 3.365 | $e^2 t_2^3$ | 4.195 |
| RhFe$_2$O$_4$ | 0.550 | 1.306 | $e^4 t_2^3$ | 6.974 |
| PdFe$_2$O$_4$ | 0.603 | 1.059 | $e^4 t_2^4$ | 7.809 |

The $\eta$ and the $\rho/\rho_{MAX}$ profiles of the TMFe$_2$O$_4$ compounds along the TM–O bonds are presented in Figure 3. VFe$_2$O$_4$, CrFe$_2$O$_4$, MnFe$_2$O$_4$, Fe$_3$O$_4$ and all of 4d TM containing compounds exhibit the covalent mid-bond $\eta$ peak (marked with grey arrows) indicating a deviation from ionicity. Furthermore, for NbFe$_2$O$_4$, MoFe$_2$O$_4$, TcFe$_2$O$_4$ and RuFe$_2$O$_4$, $\rho/\rho_{MAX}$ remains relatively larger than ~0.1 implying considerable electronic population occupying the hybridised covalent orbitals. Correspondingly, as presented in Table 2, for all compounds mentioned earlier, the calculated magnetic moment on the TM ion, is smaller than that of the nominal value designated for the high spin +3 oxidation state. This deviation from ionicity manifests in a reduction of TM magnetic moment by at least 1 $\mu_B$ in most compounds. The reduction in the magnetic moment is even more noticeable in 4d TM containing compounds where the deviation from nominal magnetisation is at least 1.5 $\mu_B$ approaching ~2 $\mu_B$ at the late 4d TM ions. Smaller magnetisation on the TM ions that cancels the Fe's magnetic moment translates to large magnetic moment per formula unit in those compounds with a strong covalent tendency in TM–O bond. This effect is most profound in PdFe$_2$O$_4$ with a magnetic moment of 7.809 $\mu_B$/f.u. which is larger by 3.817 $\mu_B$/f.u. than the parent Fe$_3$O$_4$ compound. The stability of ferrimagnetic alignment in PdFe$_2$O$_4$ is also as robust as that of Fe$_3$O$_4$ indicated by a comparable $\Delta E$ of 0.603 eV/f.u. Similar $\Delta E$ for PdFe$_2$O$_4$ ensures a similar $T_C$ as that of Fe$_3$O$_4$ which extends well beyond room temperature. Other compounds with noticeably large magnetic saturation are VFe$_2$O$_4$, CrFe$_2$O$_4$, NiFe$_2$O$_4$, NbFe$_2$O$_4$, MoFe$_2$O$_4$, RhFe$_2$O$_4$ out of which NiFe$_2$O$_4$, and RhFe$_2$O$_4$ are predicted to have above room temperature ferrimagnetic ordering.

One curious question that arises is if it is possible to obtain an even larger magnetic moment if a non-magnetic ion occupies the A site. To maintain long-range magnetic order, the ion occupying the tetrahedral site should be able of commuting a strong superexchange interaction, that is, having occupied d electrons. The exchange overlap integrals between O 2p orbitals and non-d or non-f orbital is, however, an order of magnitude smaller than the p-d exchange.[17] We verified this notion by examining the magnetic ordering of AlFe$_2$O$_4$ in which the d$^0$ Al occupies the tetrahedral site. We once calculated the total energy of the ferromagnetic state where all four Fe ions in a primitive cell were aligned parallel and once in an antiferromagnetic state where the spin of Fe ions of two adjacent primitive cells in a 1$a$ × 1$b$ × 2$c$ supercell were aligned antiparallel. We found that the ferromagnetic state was stable by 0.016 eV/f.u. This margin of stability is an order of magnitude smaller than that in TMFe$_2$O$_4$ compounds which implies no likelihood of room temperature ordering.

One major obstacle in realising the high magnetic saturation PdFe$_2$O$_4$ is the compound's possible instability against constituent elements and competing phases. To clarify this point, we calculated the formation enthalpy ($\Delta H$) of PdFe$_2$O$_4$ in oxygen-rich condition according to:

$\Delta H_{O-rich} = E^{Total}(\text{PdFe}_2\text{O}_4) - E^{Total}(\text{PdO}) - E^{Total}(\text{Fe}_2\text{O}_3)$,  Eq. 2

where $E^{Total}$ denotes the total DFT energy of inverse spinel PdFe$_2$O$_4$, P4$_2$/$mmc$ PdO and 2/$m$($C$) monoclinic Fe$_2$O$_3$. We found that $\Delta H$ was −0.692 eV/f.u. PdFe$_2$O$_4$'s $\Delta H$ in oxygen-poor condition calculated against the constituent elements according to:

$\Delta H_{O-poor} = E^{Total}(\text{PdFe}_2\text{O}_4) - E^{Total}(\text{Pd}) - 2E^{Total}(\text{Fe}) - 2E^{Total}(\text{O}_2)$,  Eq. 3

was also found substantially stable at −4.838 eV/f.u. Although PdFe$_2$O$_4$ is stable against the competing elemental and oxide constituents, we nonetheless found that another polymorph namely Fe(FePd)O$_4$ in which the Pd ions occupy the octahedral sites instead of the tetrahedral sites was more stable by 1.037 eV/f.u. Consequently, the successful synthesis of PdFe$_2$O$_4$ requires the use of non-equilibrium techniques preferably in low dimension geometry such as metal-assisted thin film crystallisation as utilised for growing CoFe$_2$O$_4$.[18] One advantageous factor in growing PdFe$_2$O$_4$ is that it crystallises in perfect $Fd\bar{3}m$ structures while Fe(PdFe)O$_4$ crystallises in a body centred orthorhombic structure with space group $Imma$ (#74) with $a = 6.059$ Å, $b = 6.140$ Å and $c = 8.631$ Å. Accordingly, any non-equilibrium PdFe$_2$O$_4$ growth can, rely on the significant difference in the lattice parameters of $Fd\bar{3}m$ PdFe$_2$O$_4$ and $Imma$ Fe(PdFe)O$_4$.

## 4. Conclusions

In summary, we demonstrated that higher covalency, quantified by the mid-bond $\eta$ peak, results in a reduced magnetic moment for the TM ion occupy the tetrahedral site of the spinel TMFe$_2$O$_4$ compounds. This reduction was found to be more substantial for PdFe$_2$O$_4$ for which the magnetisation on Pd was found to be 1.059 $\mu_B$ which is ~2 $\mu_B$ smaller than what is expected from ionic $e^4 t_2^4$ Pd resulting in a net magnetic moment of 7.809 $\mu_B$/f.u. for the compound. NiFe$_2$O$_4$ and RhFe$_2$O$_4$ were also found large magnetic moment approximately twice as large as that of magnetite all with $T_c$ above room temperature.

## Acknowledgements

Computational resources were provided by the Integrated Materials Design Centre at the University of New South Wales, Sydney, Australia.